\newif\iflanl
\openin 1 lanlmac
\ifeof 1 \lanlfalse \else \lanltrue \fi
\closein 1
\iflanl
    \input lanlmac
\else
    \message{[lanlmac not found - use harvmac instead}
    \input harvmac
    \fi
\newif\ifhypertex
\ifx\hyperdef\UnDeFiNeD
    \hypertexfalse
    \message{[HYPERTEX MODE OFF}
    \def\hyperref#1#2#3#4{#4}
    \def\hyperdef#1#2#3#4{#4}
    \def\hypernoname{}
    \def\e@tf@ur#1{}
    \def\eprt#1{{\tt #1}}
\else
    \hypertextrue
    \message{[HYPERTEX MODE ON}
\def\eprt#1{{\tt
#1}}
\fi
\newif\ifdraft

\noblackbox
\catcode`\@=11
\newif\iffrontpage
\ifx\answ\bigans
\def\titleft{\titla}
\magnification=1200\baselineskip=14pt plus 2pt minus 1pt
%
\advance\hoffset by-0.075truein
\advance\voffset by1.truecm
\hsize=6.15truein\vsize=600.truept\hsbody=\hsize\hstitle=\hsize
\else\let\lr=L
\def\titleft{\titla}
\magnification=1000\baselineskip=14pt plus 2pt minus 1pt
%
\hoffset=-0.75truein\voffset=-.0truein
\vsize=6.5truein
\hstitle=8.truein\hsbody=4.75truein
\fullhsize=10truein\hsize=\hsbody
\fi
\parskip=4pt plus 15pt minus 1pt
%
\newif\iffigureexists
\newif\ifepsfloaded
\def\epsfcheck{
\ifdraft
\input epsf\epsfloadedtrue
\else
  \openin 1 epsf
  \ifeof 1 \epsfloadedfalse \else \epsfloadedtrue \fi
  \closein 1
  \ifepsfloaded
    \input epsf
  \else
\immediate\write20{NO EPSF FILE --- FIGURES WILL BE IGNORED}
  \fi
\fi
\def\epsfcheck{}}
\def\checkex#1{
\ifdraft
\figureexistsfalse\immediate%
\write20{Draftmode: figure #1 not included}
\figureexiststrue
\else\relax
    \ifepsfloaded \openin 1 #1
        \ifeof 1
           \figureexistsfalse
  \immediate\write20{FIGURE FILE #1 NOT FOUND}
        \else \figureexiststrue
        \fi \closein 1
    \else \figureexistsfalse
    \fi
\fi}
\def\missbox#1#2{$\vcenter{\hrule
\hbox{\vrule height#1\kern1.truein
\raise.5truein\hbox{#2} \kern1.truein \vrule} \hrule}$}
\def\lfig#1{
\let\labelflag=#1%
\def\numb@rone{#1}%
\ifx\labelflag\UnDeFiNeD%
{\xdef#1{\the\figno}%
\writedef{#1\leftbracket{\the\figno}}%
\global\advance\figno by1%
}\fi{\hyperref{}{figure}{{\numb@rone}}{Fig.{\numb@rone}}}}
\def\figinsert#1#2#3#4{
\epsfcheck\checkex{#4}%
\def\figsize{#3}%
\let\flag=#1\ifx\flag\UnDeFiNeD
{\xdef#1{\the\figno}%
\writedef{#1\leftbracket{\the\figno}}%
\global\advance\figno by1%
}\fi
\goodbreak\midinsert%
\iffigureexists
\centerline{\epsfysize\figsize\epsfbox{#4}}%
\else%
\vskip.05truein
  \ifepsfloaded
  \ifdraft
  \centerline{\missbox\figsize{Draftmode: #4 not included}}%
  \else
  \centerline{\missbox\figsize{#4 not found}}
  \fi
  \else
  \centerline{\missbox\figsize{epsf.tex not found}}
  \fi
\vskip.05truein
\fi%
{\smallskip%
\leftskip 4pc \rightskip 4pc%
\noindent\ninepoint\sl \baselineskip=11pt%
{\bf{\hyperdef\hypernoname{figure}{{#1}}{Fig.{#1}}}:~}#2%
\smallskip}\bigskip\endinsert%
}

\def\boxit#1{\vbox{\hrule\hbox{\vrule\kern8pt
\vbox{\hbox{\kern8pt}\hbox{\vbox{#1}}\hbox{\kern8pt}}
\kern8pt\vrule}\hrule}}
\def\mathboxit#1{\vbox{\hrule\hbox{\vrule\kern8pt\vbox{\kern8pt
\hbox{$\displaystyle #1$}\kern8pt}\kern8pt\vrule}\hrule}}
%
\font\bigit=cmti10 scaled \magstep1

\font\titla=cmr10 scaled\magstep3

\font\absmss=cmss10 scaled\magstep1

\newfam\mssfam
\font\footrm=cmr8  \font\footrms=cmr5
\font\footrmss=cmr5   \font\footi=cmmi8
\font\footis=cmmi5   \font\footiss=cmmi5
\font\footsy=cmsy8   \font\footsys=cmsy5
\font\footsyss=cmsy5   \font\footbf=cmbx8
\font\footmss=cmss8
\def\footfont{\def\rm{\fam0\footrm}
\textfont0=\footrm \scriptfont0=\footrms
\scriptscriptfont0=\footrmss
\textfont1=\footi \scriptfont1=\footis
\scriptscriptfont1=\footiss
\textfont2=\footsy \scriptfont2=\footsys
\scriptscriptfont2=\footsyss
\textfont\itfam=\footi \def\it{\fam\itfam\footi}
\textfont\mssfam=\footmss \def\mss{\fam\mssfam\footmss}
\textfont\bffam=\footbf \def\bf{\fam\bffam\footbf} \rm}
\ifx\answ\bigans\def\abstractfont{\tenpoint}\else
\def\abstractfont{\def\rm{\fam0\absrm}
\textfont0=\absrm \scriptfont0=\absrms
\scriptscriptfont0=\absrmss
\textfont1=\absi \scriptfont1=\absis
\scriptscriptfont1=\absiss
\textfont2=\abssy \scriptfont2=\abssys
\scriptscriptfont2=\abssyss
\textfont\itfam=\bigit \def\it{\fam\itfam\bigit}
\textfont\mssfam=\absmss \def\mss{\fam\mssfam\absmss}
\textfont\bffam=\absbf \def\bf{\fam\bffam\absbf}\rm}\fi
%
\def\f@@t{\baselineskip10pt\lineskip0pt\lineskiplimit0pt
\bgroup\aftergroup\@foot\let\next}
\setbox\strutbox=\hbox{\vrule height 8.pt depth 3.5pt width\z@}
\def\vfootnote#1{\insert\footins\bgroup
\baselineskip10pt\footfont
\interlinepenalty=\interfootnotelinepenalty
\floatingpenalty=20000
\splittopskip=\ht\strutbox \boxmaxdepth=\dp\strutbox
\leftskip=24pt \rightskip=\z@skip
\parindent=12pt \parfillskip=0pt plus 1fil
\spaceskip=\z@skip \xspaceskip=\z@skip
\Textindent{$#1$}\footstrut\futurelet\next\fo@t}
\def\Textindent#1{\noindent\llap{#1\enspace}\ignorespaces}
\def\foot{\global\advance\ftno by1%
\attach{\hyperref{}{footnote}{\the\ftno}{\footsymbolgen}}%
\vfootnote{\hyperdef\hypernoname{footnote}{\the\ftno}{\footsymbol}}}%
\def\footnote#1{\global\advance\ftno by1%
\attach{\hyperref{}{footnote}{\the\ftno}{#1}}%
\vfootnote{\hyperdef\hypernoname{footnote}{\the\ftno}{#1}}}%
\newcount\lastf@@t           \lastf@@t=-1
\newcount\footsymbolcount    \footsymbolcount=0
\global\newcount\ftno \global\ftno=0
\def\footsymbolgen{\relax\footsym
\global\lastf@@t=\pageno\footsymbol}
\def\footsym{\ifnum\footsymbolcount<0
\global\footsymbolcount=0\fi
{\iffrontpage \else \advance\lastf@@t by 1 \fi
\ifnum\lastf@@t<\pageno \global\footsymbolcount=0
\else \global\advance\footsymbolcount by 1 \fi }
\ifcase\footsymbolcount
\fd@f\dagger\or \fd@f\diamond\or \fd@f\ddagger\or
\fd@f\natural\or \fd@f\ast\or \fd@f\bullet\or
\fd@f\star\or \fd@f\nabla\else \fd@f\dagger
\global\footsymbolcount=0 \fi }
\def\fd@f#1{\xdef\footsymbol{#1}}
\def\space@ver#1{\let\@sf=\empty \ifmmode #1\else \ifhmode
\edef\@sf{\spacefactor=\the\spacefactor}
\unskip${}#1$\relax\fi\fi}
\def\attach#1{\space@ver{\strut^{\mkern 2mu #1}}\@sf}
%
\newif\ifnref
\def\rrr#1#2{\relax\ifnref\nref#1{#2}\else\ref#1{#2}\fi}
\def\ldf#1#2{\begingroup\obeylines
\gdef#1{\rrr{#1}{#2}}\endgroup\unskip}

\nreffalse

\def\lref{\ldf}

\def\eqn#1{\xdef #1{(\noexpand\hyperref{}%
{equation}{\secsym\the\meqno}%
{\secsym\the\meqno})}\eqno(\hyperdef\hypernoname{equation}%
{\secsym\the\meqno}{\secsym\the\meqno})\eqlabeL#1%
\writedef{#1\leftbracket#1}\global\advance\meqno by1}
\def\eqnalign#1{\xdef #1{\noexpand\hyperref{}{equation}%
{\secsym\the\meqno}{(\secsym\the\meqno)}}%
\writedef{#1\leftbracket#1}%
\hyperdef\hypernoname{equation}%
{\secsym\the\meqno}{\e@tf@ur#1}\eqlabeL{#1}%
\global\advance\meqno by1}
\def\eqnalign#1{\xdef #1{(\secsym\the\meqno)}
\writedef{#1\leftbracket#1}%
\global\advance\meqno by1 #1\eqlabeL{#1}}
%

%
\def\chap#1{\newsec{#1}}
\def\chapter#1{\chap{#1}}
\def\sect#1{\subsec{#1}}
\def\section#1{\sect{#1}}
\def\\{\ifnum\lastpenalty=-10000\relax
\else\hfil\penalty-10000\fi\ignorespaces}
\def\note#1{\leavevmode%
\edef\@@marginsf{\spacefactor=\the\spacefactor\relax}%
\ifdraft\strut\vadjust{%
\hbox to0pt{\hskip\hsize%
\ifx\answ\bigans\hskip.1in\else\hskip .1in\fi%
\vbox to0pt{\vskip-\dp
\strutbox\sevenbf\baselineskip=8pt plus 1pt minus 1pt%
\ifx\answ\bigans\hsize=.7in\else\hsize=.35in\fi%
\tolerance=5000 \hbadness=5000%
\leftskip=0pt \rightskip=0pt \everypar={}%
\raggedright\parskip=0pt \parindent=0pt%
\vskip-\ht\strutbox\noindent\strut#1\par%
\vss}\hss}}\fi\@@marginsf\kern-.01cm}
\def\titlepage{%
\frontpagetrue\nopagenumbers\abstractfont%
\hsize=\hstitle\rightline{\vbox{\baselineskip=10pt%
{\abstractfont\pubnum}}}\pageno=0}
\frontpagefalse
\def\pubnum{}
\def\pdate{\number\month/\number\yearltd}
\def\makefootline{\iffrontpage\vskip .27truein
\line{\the\footline}
\vskip -.1truein\leftline{\vbox{\baselineskip=10pt%
{\abstractfont\pdate}}}
\else\vskip.5cm\line{\hss \tenrm $-$ \folio\ $-$ \hss}\fi}
\def\title#1{\vskip .7truecm\titlestyle{\titleft #1}}
\def\titlestyle#1{\par\begingroup \interlinepenalty=9999
\leftskip=0.02\hsize plus 0.23\hsize minus 0.02\hsize
\rightskip=\leftskip \parfillskip=0pt
\hyphenpenalty=9000 \exhyphenpenalty=9000
\tolerance=9999 \pretolerance=9000
\spaceskip=0.333em \xspaceskip=0.5em
\noindent #1\par\endgroup }
\def\autskip{\ifx\answ\bigans\vskip.5truecm\else\vskip.1cm\fi}
\def\author#1{\vskip .7in \centerline{#1}}

\def\address#1{\ifx\answ\bigans\vskip.2truecm
\else\vskip.1cm\fi{\it \centerline{#1}}}
\def\abstract#1{
\vskip .5in\vfil\centerline
{\bf Abstract}\penalty1000
{{\smallskip\ifx\answ\bigans\leftskip 2pc \rightskip 2pc
\else\leftskip 5pc \rightskip 5pc\fi
\noindent\abstractfont \baselineskip=12pt
{#1} \smallskip}}
\penalty-1000}
\def\ack{\vskip1.6cm\centerline{{\bf Acknowledgements}}}
%
%

%
\def\bfone{\relax{\rm 1\kern-.35em 1}}
\def\inbar{\vrule height1.5ex width.4pt depth0pt}
\def\IC{\relax\,\hbox{$\inbar\kern-.3em{\mss C}$}}
\def\ID{\relax{\rm I\kern-.18em D}}
\def\IF{\relax{\rm I\kern-.18em F}}
\def\IH{\relax{\rm I\kern-.18em H}}
\def\II{\relax{\rm I\kern-.17em I}}
\def\IN{\relax{\rm I\kern-.18em N}}
\def\IP{\relax{\rm I\kern-.18em P}}
\def\IQ{\relax\,\hbox{$\inbar\kern-.3em{\rm Q}$}}
\def\IR{\relax{\rm I\kern-.18em R}}
\def\ZZ{\relax{\hbox{\mss Z\kern-.42em Z}}}

\def\nup#1({Nucl.\ Phys.\ $\us {B#1}$\ (}
\def\plt#1({Phys.\ Lett.\ $\us  {#1}$\ (}
\def\cmp#1({Comm.\ Math.\ Phys.\ $\us  {#1}$\ (}
\def\prp#1({Phys.\ Rep.\ $\us  {#1}$\ (}
\def\prl#1({Phys.\ Rev.\ Lett.\ $\us  {#1}$\ (}
\def\prv#1({Phys.\ Rev.\ $\us  {#1}$\ (}
\def\mpl#1({Mod.\ Phys.\ Let.\ $\us  {A#1}$\ (}
\def\ijmp#1({Int.\ J.\ Mod.\ Phys.\ $\us{A#1}$\ (}
\def\tit#1|{{\it #1},\ }
%

%

\def\bar{\overline}
\def\us#1{\underline{#1}}

\def\hat{\widehat}

\def\Coe#1.#2.{{#1\over #2}}

\def\coe#1.#2.{\relax{\textstyle {#1 \over #2}}\displaystyle}

\def\notin{\hbox{{$\in$}\kern-.51em\hbox{/}}}

\def\del{\partial}

\def\eg{{\it e.g.}}
\def\ie{{\it i.e.}}

\catcode`\@=12



\def\ijmp {{Int. J. Mod. Phys.\ }{\bf A}}

\lref\willy{
S.~L.~Shatashvili and C.~Vafa,
``Superstrings and manifolds of exceptional holonomy,''
arXiv:hep-th/9407025.
}

\lref\gorilla{
J.~M.~Figueroa-O'Farrill,
``A note on the extended superconformal algebras associated with  
manifolds of exceptional holonomy,''
Phys.\ Lett.\ B {\bf 392}, 77 (1997)
[arXiv:hep-th/9609113].
}

\lref\GENO{
D.~Gepner and B.~Noyvert,
``Unitary representations of SW(3/2,2) superconformal algebra,''
Nucl.\ Phys.\ B {\bf 610}, 545 (2001)
[arXiv:hep-th/0101116].
}

\lref\EGSU{
T.~Eguchi and Y.~Sugawara,
``CFT description of string theory compactified on non-compact manifolds
with G(2) holonomy,''
Phys.\ Lett.\ B {\bf 519}, 149 (2001)
[arXiv:hep-th/0108091].
}

\lref\SUYA{
K.~Sugiyama and S.~Yamaguchi,
``Cascade of special holonomy manifolds and heterotic string theory,''
arXiv:hep-th/0108219.
}

\lref\PAPI{
H.~Partouche and B.~Pioline,
``Rolling among G(2) vacua,''
JHEP {\bf 0103}, 005 (2001)
[arXiv:hep-th/0011130].
}

\lref\GRGU{
M.~B.~Green and M.~Gutperle,
``Light-cone supersymmetry and D-branes,''
Nucl.\ Phys.\ B {\bf 476}, 484 (1996)
[arXiv:hep-th/9604091].
}

\lref\GUSA{
M.~Gutperle and Y.~Satoh,
``D-branes in Gepner models and supersymmetry,''
Nucl.\ Phys.\ B {\bf 543}, 73 (1999)
[arXiv:hep-th/9808080].
}

\lref\FUSW{
J.~Fuchs, C.~Schweigert and J.~Walcher,
``Projections in string theory and boundary states for Gepner models,''
Nucl.\ Phys.\ B {\bf 588}, 110 (2000)
[arXiv:hep-th/0003298].
}

\lref\CARD{
J.~L.~Cardy,
``Boundary Conditions, Fusion Rules And The Verlinde Formula,''
Nucl.\ Phys.\ B {\bf 324}, 581 (1989).
}

\lref\OOOY{
H.~Ooguri, Y.~Oz and Z.~Yin,
``D-branes on Calabi-Yau spaces and their mirrors,''
Nucl.\ Phys.\ B {\bf 477}, 407 (1996)
[arXiv:hep-th/9606112].
}

\lref\BBMOOY{
K.~Becker, M.~Becker, D.~R.~Morrison, H.~Ooguri, Y.~Oz and Z.~Yin,
``Supersymmetric cycles in exceptional holonomy manifolds and Calabi-Yau  4-folds,''
Nucl.\ Phys.\ B {\bf 480}, 225 (1996)
[arXiv:hep-th/9608116].
}

\lref\BDLR{
I.~Brunner, M.R.~Douglas, A.~Lawrence and C.~R\"omelsberger,
``D-branes on the quintic,''
\eprt{hep-th/9906200}. 
}

\lref\DOFR{
M.~R.~Douglas, B.~Fiol and C.~Romelsberger,
``Stability and BPS branes,''
arXiv:hep-th/0002037.
}

\lref\DOUG{
M.~R.~Douglas,
``D-branes, categories and N = 1 supersymmetry,''
arXiv:hep-th/0011017.
}

\lref\SCYA{
A.~N.~Schellekens and S.~Yankielowicz,
``Simple Currents, Modular Invariants And Fixed Points,''
Int.\ J.\ Mod.\ Phys.\ A {\bf 5}, 2903 (1990).
}

\lref\GEPN{
D.~Gepner,
``Space-Time Supersymmetry In Compactified String Theory And Superconformal Models,''
Nucl.\ Phys.\ B {\bf 296}, 757 (1988).
}

\lref\FKSS{
J.~Fuchs, A.~Klemm, C.~Scheich and M.~G.~Schmidt,
``Spectra And Symmetries Of Gepner Models Compared To Calabi-Yau Compactifications,''
Annals Phys.\  {\bf 204}, 1 (1990).
}

\lref\LYSC{
M.~Lynker and R.~Schimmrigk,
``A-D-E Quantum Calabi-Yau Manifolds,''
Nucl.\ Phys.\ B {\bf 339}, 121 (1990).
}

\lref\DVVV{
R.~Dijkgraaf, C.~Vafa, E.~Verlinde and H.~Verlinde,
``The Operator Algebra Of Orbifold Models,''
Commun.\ Math.\ Phys.\  {\bf 123}, 485 (1989).
}

\lref\BIFS{
L.~Birke, J.~Fuchs and C.~Schweigert,
``Symmetry breaking boundary conditions and WZW orbifolds,''
Adv.\ Theor.\ Math.\ Phys.\  {\bf 3}, 671 (1999)
[arXiv:hep-th/9905038].
}

\lref\GANN{
T.~Gannon,
``Towards a classification of SU(2) x ... x SU(2) modular invariant partition functions,''
J.\ Math.\ Phys.\  {\bf 36}, 675 (1995)
[arXiv:hep-th/9402074].
}

\lref\FUSC{
J.~Fuchs and C.~Schweigert,
``Category theory for conformal boundary conditions,''
arXiv:math.ct/0106050.
}

\lref\HITC{
N.~Hitchin,
``The geometry of three-forms in six and seven dimensions,''
arXiv:math.dg/0010054.
}

\lref\HOVA{
K.~Hori and C.~Vafa,
``Mirror symmetry,''
arXiv:hep-th/0002222.
}

\lref\TALK{
J.~Walcher, 
``Rational Conformal Field Theories with $G_2$ Holonomy,''
talk at ITP Santa Barbara, and CIT-USC  CThPh, October 2001, 
available online at
{\tt http://www.itp.ucsb.edu/online/joint98/walcher/}
}

\lref\BB{
R.~Blumenhagen and V.~Braun,
``Superconformal Field Theories for Compact $G_2$ Manifolds,''
arXiv:hep-th/0110232.
}

\lref\ja{D.~Joyce, 
``Compact Riemannian 7-manifolds with holonomy G2. I'',
J. Diff. Geom. {\bf 43}  (1996)  291. 
}

\lref\jb{D.~Joyce, 
``Compact Riemannian 7-manifolds with holonomy G2. II'', 
J. Diff. Geom. {\bf 43} (1996) 329 
}

\lref\BR{R.L.~Bryant, ``Metrics with Exceptional Holonomy'', 
Ann.Math. {\bf 126} (1987) 525
}

\lref\BRSA{R.L.~Bryant and S.M.~Salamon, ``On the Construction of 
some complete metrics with exceptional holonomy'', Duke Math. J. {\bf 58} (1989) 829
}

\lref\GR{A.~Gray, ``Weak Holonomy Groups'', Math. Z. {\bf 123} (1971) 290}

\lref\BEJO{M.~Bershadsky, A.~Johansen, ``Large N limit of orbifold field theories'', Nucl.Phys. 
{\bf B536} (1998) 141}


\font\mybbb=msbm10 at 8pt
\font\mybb=msbm10 at 10pt
\font\myfrak=eufm10 at 10pt
\def\bbb#1{\hbox{\mybbb#1}}
\def\bb#1{\hbox{\mybb#1}}
\def\frak#1{\hbox{\myfrak#1}}

\def\Re {\bb{R}}
\def\Z {\bb{Z}}
\def\sZ{\bbb{Z}}
\def\P {\bb{P}}

\def\fus{\star}
\def\frac#1#2{{#1\over #2}}

\def\ZZ{{\bb{Z}}}

\def\calc{{{\cal C}}}
\def\calo{{{\cal O}}}

\def\ee{{\rm e}}
\def\ii{{\rm i}}
\def\text#1{{\rm #1 }}

\def\N{{\cal N}}

\def\IG{\relax\,\hbox{$\inbar\kern-.3em{\mss G}$}}

\def\nocite{\phantom}

\def\id{\protect{{1 \kern-.28em {\rm l}}}}


\def\cite{}



\def\pubnum{
\hbox{NSF-ITP-01-166}
\hbox{hep-th/0110302}
\hbox{}}
\def\pdate{}
\titlepage
\vskip-.5cm
\Title{}{Rational Conformal Field Theories With $G_2$ Holonomy}
\vskip -1.2cm
\autskip
\author{R.\ Roiban$^a$ and J.\ Walcher$^b$}
\vskip0.7truecm
\it
\centerline{${}^a$Department of Physics}
\centerline{${}^b$Institute for Theoretical Physics}
\vskip 0.2cm
\centerline{University of California}
\centerline{Santa Barbara, CA 93106, U.S.A.}
\rm
\abstract{%
We study conformal field theories for strings propagating on compact, 
seven-dimensional manifolds with $G_2$ holonomy. In particular, we describe
the construction of rational examples of such models. We argue that analogues
of Gepner models are to be constructed based not on $\N=1$ minimal models, but on
$\ZZ_2$ orbifolds of $\N=2$ models. In $\ZZ_2$ orbifolds of Gepner models times a
circle, it turns out that unless all levels are even, there are no new Ramond ground
states from twisted sectors. In examples such as the quintic Calabi-Yau, this reflects
the fact that the classical geometric orbifold singularity can not be resolved without
violating $G_2$ holonomy. We also comment on supersymmetric boundary states in such
theories, which correspond to D-branes wrapping supersymmetric cycles in the
geometry.}

\Date{\vbox{\hbox{ October 2001}}}
\goodbreak

\parskip=4pt plus 15pt minus 1pt
\baselineskip=14pt 
\leftskip=8pt \rightskip=10pt
%


\newsec{Introduction}

Supersymmetric compactifications of higher-dimensional theories require the
internal part of spacetime to have special properties. In the context of
supergravity, string, and M-theory, the classical starting points are manifolds
of special holonomy, such as Calabi-Yau manifolds, which are $2n$-dimensional
manifolds with ${\rm SU}(n)\subset {\rm SO}(2n)$ holonomy. Additional insight
can be gained by probing the geometry with the various extended objects
provided by string/M-theory. The goal of the present paper is to explore the
geometry of compact seven-dimensional Riemannian manifolds with exceptional 
holonomy group $G_2$, using strings as probes. Such manifolds are required for 
minimal supersymmetry when compactifying $7=11-4=10-3$ dimensions.
\nocite{\willy}
\nocite{\gorilla}

The subject of string theory on manifolds with $G_2$ holonomy was started
in \willy, where the superconformal algebra that is expected to characterize
$G_2$ holonomy was derived. It turns out that the r\^ole played by the
rational ${\rm U}(1)$ current algebra in the Calabi-Yau context is here
played by the tri-critical Ising model.

Let us briefly review how this comes about. For a generic simply connected
$d$-dimensional Riemannian manifold $M$, the holonomy group is the Lie group
${\rm SO}(d)$. Also, the worldsheet fermions of a supersymmetric $\sigma$-model on 
$M$ come in a representation of (local) ${\rm SO}(d)$. If $M$ is flat and the holonomy
trivial, the fermions are free and give rise to an $\frak{so}(d)_1$ current algebra 
on the worldsheet. If the holonomy is a non-trivial special subgroup $G\subset 
{\rm SO}(d)$, the worldsheet fermions interact through the coupling to the connection 
on $TM$ which takes values in the Lie algebra $\frak{g}$ of $G$. To see what CFT might
describe these fermions, recall that gauged WZW models provide a Lagrangian description 
of coset models of conformal field theories. By analogy, the residual symmetry 
associated with special holonomy $G$ is expected to be the symmetry algebra of the
coset CFT,
$$
\frac{{\frak{so}}(d)_1}{\frak{g}}\,.
\eqn\coset
$$
For instance, for $d=2n$ even, the special holonomy ${\rm SU}(n)$ leads to the coset
CFT ${\frak{so}}(2n)_1/{\frak{su}}(n)$, which is nothing but a $\frak{u}(1)$ current
algebra extended by a spin $2n=2\hat{c}$ field related to the spectral flow
operator. If $d=7$, the exceptional holonomy group $G_2\subset {\rm SO}(7)$ leads 
to the coset $\frak{so}(7)_1/G_2$, which turns out to have central charge $7/10$
and hence is the CFT of the Ising model at the tri-critical point \willy.
\nocite{\refs{\GENO,\SUYA,\EGSU,\PAPI,\GEPN,\FKSS,\LYSC,\SCYA,\GANN}}
\nocite{\refs{\GRGU,\OOOY,\GUSA,\FUSW,\DOFR,\DOUG,\CARD,\BBMOOY}}
\nocite{\refs{\DVVV,\BIFS,\FUSC}}

Shatashvili and Vafa \willy\ also derive the extension of the $\N=1$ superconformal
algebra by the tri-critical Ising algebra from a free field representation. In this 
approach, the spin $3/2$ field of the tri-critical Ising model arises from the closed 
three-form $\phi$ that determines the $G_2$ structure. The stress tensor at central 
charge $7/10$ is essentially the dual, closed $4$-form $*\phi$.
 
In \gorilla, the superconformal algebra associated with $G_2$ holonomy was 
rederived by representing it as the fixed point algebra of the Calabi-Yau
three-fold algebra times $\frak{u}(1)$ under the $\ZZ_2$ automorphism which acts as
the mirror automorphism on the $\N=2$ algebra and as the usual $\ZZ_2$ 
automorphism on $\frak{u}(1)$. This mimics the general proposal of Joyce for the 
construction of $G_2$ holonomy manifolds, which involves orbifolding a Calabi-Yau 
manifold times a circle by a $\ZZ_2$ which acts as an anti-holomorphic involution
on the Calabi-Yau and inversion on the circle. 

Abstracting the algebra from \refs{\willy,\gorilla}, it becomes a natural
problem to look for explicit realizations. Recent work on the subject includes the
study of the representation theory of closely related ${{\cal W}}$-algebras \GENO,
the relation between the algebras associated with various special holonomies
\SUYA, as well as the construction of modular invariant partition functions for strings
on non-compact manifolds with $G_2$ holonomy \EGSU.
\nocite{\refs{\HOVA,\BDLR}}
\nocite{\refs{\ja,\jb,\GR,\BR,\BRSA,\HITC,\BEJO}}

One of the purposes of the present work is to start the construction of explicit
examples of compact, rational, conformal field theories with $G_2$ holonomy.
As is well-appreciated in the context of Calabi-Yau manifolds, the study of conformal
field theories and their chiral algebras is only one end of the spectrum of approaches
to exploring (even perturbative) string theory. A substantial amount of information can
already be gained from classical geometry, or from the low-energy space-time theory,
as well as from effective worldsheet descriptions such as Landau-Ginzburg models or
gauged linear $\sigma$-models. How much of this will be available for manifolds of
$G_2$ holonomy remains to be seen. However, exactly solvable models provide structural
information and are certainly a welcome starting point, for instance for a better
understanding of mirror symmetry for $G_2$ holonomy manifolds \willy.

The most readily accessible class of examples are $\ZZ_2$ orbifolds of Gepner models
times a free boson and fermion. The conformal field theory analysis in section 3 of the
present paper reveals that the twisted sector generically does not contribute any new
Ramond ground states. The only exception are the cases where all levels of the $\N=2$
minimal models are even. In the geometric description, the corresponding orbifolds of
Calabi-Yaus times a circle are singular manifolds with $G_2$ holonomy. It appears
that the orbifold can be smoothened to a $G_2$ holonomy manifold only if the first
Betti number of the fixed point set on the Calabi-Yau is non-zero. This is discussed,
for instance, in ref.\ \PAPI. For the quintic Calabi-Yau, we will verify that this
agrees with the conformal field theoretic prediction of absence of twisted sector
ground states.

An obvious question one should ask is how generic is the structure unveiled in
Gepner models for $G_2$ holonomy manifolds of physical interest, and whether one
can learn anything interesting about M-theory compactifications? Our results,
together with the failure of attempts to construct simpler models based, \eg, on 
tensor products of $\N=1$ minimal models, seem to indicate that either the best rational
models of $G_2$ holonomy are yet to be found, or that smooth $G_2$ holonomy manifolds
generically do not have a Gepner point. If, on the other hand, the structure found
in $\ZZ_2$ orbifolds of Gepner models turns out to be generic, it should imply
that many of the tools used to study Calabi-Yaus manifolds carry over, albeit with 
severe technical complications, to $G_2$ holonomy.

\noindent
{\it Note added:} The results discussed in this paper were first presented
in \TALK. While writing up this note, the preprint \BB\ was received on the
archive. In independent work, the authors of \BB\ analyze three examples
of $\ZZ_2$ orbifolds of Gepner models. Their conformal field theory results 
are in agreement with our general formulae.

\newsec{$G_2$ holonomy CFT}

The goal of this section is to develop some general ideas about the
construction of RCFTs with $G_2$ holonomy, guided by the success of
Gepner's construction of models which are exact solutions of
$\sigma$-models in terms of Calabi-Yau manifolds.

\subsec{Rational conformal field theories with special holonomy}

It is well-known that given any rational conformal field with $\N=2$
supersymmetry and total central charge $\hat{c}=\frac c3$ integer, one can
obtain the internal sector of a supersymmetric string compactification to
$D=10-2\hat c$ dimensions by projecting the theory onto integral ${\rm U}(1)$
charge. Namely, integrality of the ${\rm U}(1)$ charge is equivalent to
locality of the chiral spectral flow operator, which upon GSO projection
yields a supersymmetry in spacetime. The criterion for consistency
of the projection is here modular invariance of the torus partition function.

The most popular class of examples are Gepner's models \GEPN, in which the
starting point is a tensor product of $\N=2$ minimal models, which can be
thought of as cosets $\frac{{\rm SU}(2)_k\times {\rm U}(1)}{{\rm U}(1)}$. It
turns out that Gepner models are related to the exact solution of $\sigma$-models
on Calabi-Yau manifolds describable as complete intersections in products of 
projective spaces (see \refs{\FKSS,\LYSC} for a summary). In principle, this 
construction works for Calabi-Yau manifolds of any dimension, although from 
a physical point of view, $n=\hat c=3$ is of course the most interesting one.

Given the success of Gepner models in describing manifolds of special
holonomy ${\rm SU}(n)\subset {\rm SO}(2n)$, it is natural to wonder whether
there exist similar constructions also for the exceptional holonomy groups
of Riemannian manifolds, \ie, $G_2$ in seven dimensions, and ${\rm Spin}(7)$ 
in eight dimensions.

\subsec{Tensor products of $\N=1$ minimal models?}

For manifolds with $G_2$ holonomy, there should only be $\N=1$ supersymmetry on
the worldsheet. One might therefore be tempted to generalize Gepner's
construction by replacing $\N=2$ minimal models with $\N=1$ superconformal
minimal models, and to look for a modular invariant that projects out all
unwanted states and contains the tri-critical Ising CFT in its maximally extended
chiral algebra. It is easy to see that this construction, if it works at all, must
be very special.

First of all, it is well-known that from the series of $\N=1$ minimal models,
which are the cosets $\frac{{\rm SU}(2)_k\times {\rm SU}(2)_2}{{\rm SU}(2)_{k+2}}$, 
only those with even $k$ possess a Ramond ground state, while those with odd $k$ 
break supersymmetry spontaneously. Thus, in the list of tensor products of $\N=1$ 
minimal models with total central charge
$c=\sum c_i=\sum \left(\frac 32-\frac{12}{(k_i+2)(k_i+4)}\right)=21/2$, one has to
restrict to those were all $k_i$ are even\foot{In particular, this excludes the
tri-critical Ising model as an elementary building block.}. Then, the only candidate
for the space-time supercharge (the analog of the spectral flow operator),
with tri-critical Ising dimension $7/16$, is built out of the product of Ramond
ground states in the individual minimal models. This field however, with
minimal model labels $(k/2,k/2+1,1)$, has the fusion rules
$$
(k/2,k/2+1,1)\fus (k/2,k/2+1,1) = \sum_{l,m,s \;{\rm even}
\atop \frac l2+ \frac m2 + \frac s2 \;{\rm even}} (l,m,s) \,.
\eqn\fusion
$$
Thus, to reproduce the tri-critical Ising fusion rules, 
$\left[\frac 7{16}\right]\fus \left[\frac 7{16}\right]=[0]+\left[\frac 32\right]$, 
in the projected tensor product the modular invariant has to be such that most
of the terms on the r.h.s.\ of eq.\ \fusion\ are projected out. In particular,
this is not possible with an ordinary simple-current modular invariant. The modular
invariant has to be {\it exceptional}. In terms of modular data, one is looking 
for a modular invariant in tensor products of ${\rm SU}(2)$ WZW models with
many factors. While a number of exceptional modular invariants for these theories
are known \GANN, none of them appears to have the desired properties. A brief
explicit search in the list of candidate tensor products has not revealed any magic
in the modular transformations of the products of minimal model characters.

\subsec{Orbifolds of $\N=2$ models}

Another possibility for the construction of $G_2$ holonomy RCFTs is to combine
ordinary Gepner models with the results of ref.\ \gorilla. Namely, according to
\gorilla, any conformal field theory constructed as the $\ZZ_2$ orbifold theory of 
a Calabi-Yau model times a free boson and fermion has $G_2$ holonomy, provided the 
$\ZZ_2$ acts as 
$$
\omega: \qquad
\eqalign{
T_{CY} &\mapsto T_{CY} \;\;\,\quad\qquad T_{S^1} \mapsto T_{S^1} \cr
G_{CY} &\mapsto G_{CY} \;\;\,\quad\qquad j_{S^1} \mapsto -j_{S^1} \cr
\ii\del X := J_{CY} &\mapsto -J_{CY} \qquad\quad 
\psi_{S^1} \mapsto - \psi_{S^1}  \cr
\ee^{\ii\sqrt{\hat{c}} X} &\mapsto \ee^{-\ii\sqrt{\hat{c}}X}
} \qquad
\eqn\cyauto
$$
on the symmetry generators of the $CY\times S^1$. In particular, one may model the
Calabi-Yau by a Gepner model. We are then interested in the orbifold
$$
\left({\rm Gep} \times S^1\right)/\ZZ_2 \,.
$$

\subsec{Orbifolds and Extensions}

To construct such an orbifold, it is helpful to recall the following well-known fact:
The orbifold of any given theory allows for an inverse operation which returns the
original theory. In the geometrical context, if the orbifold group is abelian, the
inverse operation is often another orbifold by the same group. In the language of 
conformal field theory, were orbifolding amounts to reducing the chiral symmetry,
one recovers the original theory by extending the orbifold theory by all symmetry
generators that were broken in the orbifold construction by the introduction of twist
fields. In particular, to any $\ZZ_2$ orbifold of a given CFT $\calc$, there is a $\ZZ_2$
simple-current extension of $\calc/\ZZ_2$, such that $(\calc/\ZZ_2)^{\sZ_2}=\calc$.
We refer to ref.\ \BIFS\ for very explicit and extremely useful formulae 
on general $\ZZ_2$ orbifolds, and WZW orbifolds in particular. To construct $G_2$ 
holonomy models starting from Gepner models, we thus use the formula
$$
\left({\rm Gep}\times S^1\right)/\ZZ_2 =
\left({\rm Gep}/\ZZ_2 \times S^1/\ZZ_2\right)^{\sZ_2} \,.
\eqn\central
$$
where we extend by the off-diagonal $\Z_2$.

In turn, the $\ZZ_2$ orbifold of a Gepner model is most easily constructed by
thinking of a Gepner model as an extended tensor product of $\N=2$ minimal
models. Namely, one orbifolds the individual models, and reconstructs the orbifold
of the Gepner model by appropriately extending their tensor product. 

We would like to mention an obvious problem at this point. Before orbifolding, the
spectral flow operator which we desire to include in the chiral algebra is
a primary field of quantum dimension one, with easy to implement $\ZZ_H$ selection
rules. In fact, simple-current theory \SCYA\ provides general solutions for the
corresponding CFT problem. The problem is that after orbifolding, the generator of
the extending sector is a primary field of quantum dimension $2$, and there is no
generally known prescription for implementing the extension.

The following simple argument shows why orbifold and extension in fact do commute, at
least at the level of modular invariance. We may write the partition function of
the orbifold before extension as
$$
Z_{\rm orb} = \frac 12 \left( Z_{++} + Z_{+-} + Z_{-+} + Z_{--}\right)  \,,
$$
where the terms on the r.h.s.\ are the partition functions with twisted boundary 
conditions in space and/or time direction on the torus worldsheet. Since $Z_{++}$ 
is nothing but the original partition function, it is in particular modular invariant, 
as is the combination of the remaining three terms, $Z_{+-} + Z_{-+} + Z_{--}$. Thus, 
if we wish to construct the orbifold of an extension, we can consider the modular
invariant expression
$$
Z_{\rm orb}^{\rm ext} = \frac 12 \left( Z_{++}^{\rm ext} + Z_{+-} + Z_{-+} + 
Z_{--}\right) \,,
\eqn\orbext
$$
where $Z_{++}^{\rm ext} \equiv Z^{\rm ext}$ is the partition function of the theory
obtained by extension alone. $Z^{\rm ext}$ is modular invariant by construction. To
ensure that eq.\ \orbext\ indeed is the desired partition function, it suffices to check
that the extension does not modify the structure of twisted sectors. In other words, one
has to verify that all symmetric sectors of the theory appear in the extension
$Z^{\rm ext}$ and that the simple current does not alter the twining characters.
A simple example where this condition is satisfied is when one extends by a simple
current of odd order and the orbifold group acts by charge conjugation. But more general
results would certainly be welcome. One might expect the new conformal field theory
methods which have emerged in ref.\ \FUSC\  to be powerful enough to treat such cases.

\subsec{Supersymmetric boundary conditions and Cardy states in the
tri-critical Ising model}

The CFT of the tri-critical Ising model present in the chiral algebra associated
with $G_2$ holonomy has another important application when it comes to 
specifying possible supersymmetric boundary conditions on the worldsheet, and
is hence relevant for the geometry of D-branes wrapping supersymmetric cycles in
the $G_2$ holonomy manifold. Let us pause here from the main thread of the paper
to explain this connection.

When constructing boundary conditions in conformal field theory as a worldsheet
description of D-branes in string theory, it is important to understand the realization
of symmetries, and in particular, the conditions imposed by worldsheet and
space-time supersymmetry \refs{\GRGU,\OOOY,\GUSA,\FUSW}. More precisely, because string
theory requires gauging $\N=1$ superconformal symmetry on the worldsheet, one has
to require that the boundary conditions on the worldsheet be superconformal
invariant. For the corresponding boundary states, this imposes conditions of the
form,
$$
\eqalign{
\left( L_n - \bar{L}_{-n} \right)         |\, a \;\, \rangle\!\!\!\!\rangle &=0 \cr
\left( G_r + \ii\,\eta\, \bar{G}_{-r} \right) |\, a \;\, \rangle\!\!\!\!\rangle &=0 
\,.}
\eqn\wssusy
$$
These conditions must be satisfied exactly by the boundary states. Any additional
operators that might be present in the chiral algebra can, however, also be realized 
twistedly, \ie,
$$
\left(\calo_n - (-1)^{\Delta_{\calo}}\omega(\bar{\calo}_{-n})\right)
|\, a \;\, \rangle\!\!\!\!\rangle =0\;,
\eqn\twist
$$
where $\omega$ is some automorphism of the chiral algebra. In particular,
fields responsible for spacetime supersymmetry must satisfy a condition of
the form \twist, such that the branes preserve a combination of left- and 
right-moving spacetime supersymmetry charges, as dictated by $\omega$.

A good example for these considerations arises in Calabi-Yau compactifications.
Recall that in this case, the $\N=1$ superconformal algebra is extended by
the chiral algebra of a free boson at a rational radius, $R^2= \hat c$. Conformal
boundary conditions for the $\frak{u}(1)$ current $J=\ii\sqrt{\hat c}\,\del X$ 
are
$$
\eqalign{
{\rm Neumann} \qquad \left( J_n + \bar{J}_{-n}\right)
|\, a \;\, \rangle\!\!\!\!\rangle &= 0 \,,\cr
{\rm and\; Dirichlet} \qquad \left( J_n - \bar{J}_{-n}\right)
|\, a \;\, \rangle\!\!\!\!\rangle &= 0 \,.}
$$
Furthermore, the extension by the spectral flow operator,
$\ee^{\ii\sqrt{\hat c}\,X}$, allows the detection of the position of the Dirichlet
boundary condition or the value of the Wilson line as an automorphism type of the
boundary condition, \ie, one has the gluing condition,
$$
\ee^{\ii\sqrt{\hat c}\, X_{\rm L}(z)} =
\ee^{2\ii\varphi} \ee^{\pm\ii\sqrt{\hat c}\, X_{\rm R}(\bar z)}\,,
$$
at the worldsheet boundary $z=\bar z$. Note that $\varphi$ is well-defined only up to
shifts by $\pi$ and hence only specifies the boundary condition for $X$ modulo
$2\pi/\sqrt{\hat c}$ on a circle of radius $\sqrt{\hat c}$. This 'grade'
$\varphi$ of the boundary condition controls many essential aspects of spacetime
supersymmetry for D-branes on Calabi-Yaus, and is very important in applications.
This was particularly emphasized in ref.\ \DOUG, and further studied, \eg, in 
\refs{\DOFR,\DOUG}.

Similar reasoning is easily applied for $G_2$ holonomy, where the r\^ole of the rational
$\frak{u}(1)$ current algebra is taken over by the tri-critical Ising CFT. It is known
from the work of Cardy \CARD\ that symmetry preserving boundary conditions in rational
conformal field theories are in one-to-one correspondence to primary fields. In the
bosonic description, there are therefore six possible types of symmetric boundary 
conditions, labelled by the six primary fields of the tri-critical Ising model. In a
supersymmetric language, we expect branes and anti-branes as well as two possible
spin structures for the fermions on the open string. As regards spacetime supersymmetry,
the six primary fields of the tri-critical Ising model are accordingly divided into
two groups,
$$
\eqalign{
\left[0\right] \qquad \left[\frac 32\right] \qquad & \qquad 
\left[\frac 1{10} \right] \qquad \left[\frac 35\right] \cr
\left[\frac 7{16}\right] \;\;\;\quad\qquad & \qquad\qquad\; \left[\frac 3{80}\right]
\,.}
\eqn\boco
$$
The doubling on the first line reflects the existence of branes and anti-branes,
while the states in the second line are necessary for obtaining an open string
Ramond sector. It would be interesting to find an explicit connection
between the abstract labelling \boco\ and geometric boundary conditions at
large volume, discussed for example in \BBMOOY.

\newsec{Examples}

We now turn to the main results of this paper, which is the construction of
examples of rational conformal field theories with $G_2$ holonomy. The general
strategy, as outlined in the previous section, is to use orbifolds of $\N=2$
(minimal) models as basic building blocks, and to perform a modular invariant
projection on their tensor products. At the end of this section, we will give 
a geometric interpretation to some of our results.

\subsec{Orbifolds of $N=2$ minimal models}

We thus need $\ZZ_2$ orbifolds of $\N=2$ minimal models. Note that these orbifolds
are not the ones that are usually studied in the context of LG orbifolds, with 
action $\Phi\mapsto \ee^{2\pi\ii/h} \Phi$ on the Landau--Ginzburg field $\Phi$. In 
the CFT, the latter orbifolds arise from ``dividing out'' simple-current symmetries. 
The orbifolded theories differ from the original ones only by a modification of the 
modular invariant partition function. In particular, the symmetry algebra of the 
orbifold models still is the $\N=2$ super-Virasoro algebra.

The orbifolds of present interest arise from dividing out the mirror automorphism
of the $\N=2$ super-Virasoro algebra,
$$
\omega: \quad L_n \mapsto L_n\,, \qquad 
G^{\pm}_r \mapsto G^{\mp}_r\,, \qquad
J_n \mapsto -J_n \,.
\eqn\mimoauto
$$
In particular, the orbifold breaks $\N=2$ supersymmetry down to $\N=1$. 

The induced action on the primary fields is
$$
\omega^*: \quad \phi_{(l,m,s)} \mapsto \phi_{(l,-m,-s)}\,.
\eqn\primauto
$$
Here, $0\le l\le k$, $m\in \ZZ_{2(k+2)}$, $s\in\ZZ_4$, with $l+m+s$ even, are the
ordinary minimal model labels. Taking into account the field identification 
$(l,m,s)\equiv (k-l,m+h,s+2)$, one easily enumerates the fixed points of $\omega^*$: 
In the NS sector one has $(l,0,0)$, $(l,h,0)$, where $l=0,\ldots,k$ has the appropriate 
parity, while in the R sector, one
finds $(k/2,h/2,1)$ and $(k/2,h/2,-1)$, if $k$ is even. This gives a total of $k+1$
fixed fields if $k$ is odd, and $k+4$ if $k$ is even. By general properties of $\ZZ_2$
orbifolds \refs{\DVVV,\BIFS}, each of these fixed fields yields two primary fields in the
untwisted sector, and gives rise to two fields in the twisted sector. 

To compute the twisted sectors note that the $\N=2$ minimal models have a
representation as cosets, $\frac{{\rm SU}(2)_k\times {\rm U}(1)}{{\rm U}(1)}$.
Therefore, the orbifold of interest can be thought of as a 'coset of orbifolds'.
We note that this involves a non-standard CFT construction, and we do not
have a proof that one may interchange cosetting and orbifolding in general.
Here, however, we are dealing with a very simple case. The coset does not
have field identification fixed points, and the $\ZZ_2$ automorphism is simply
given by charge conjugation. For the ${\rm U}(1)$ factors, the classic
reference is \DVVV, while for ${\rm SU}(2)$, the relevant results can be extracted 
from \BIFS. One may then proceed formally as for the usual coset construction to 
obtain the 'orbifold of the coset' as a 'coset of orbifolds'. 

According to this presription, the twisted sectors in the orbifold of the
$\N=2$ minimal model are given by combinations $(\lambda,\mu,\sigma,\psi)$. Here,
$\lambda=0,\ldots,k$ characterizes the twisted sector in the ${\rm SU}(2)$ part,
$\mu,\sigma=0,1$ distinguish the two twisted sectors of the ${\rm U}(1)$
factors, while $\psi=\pm$ is the usual degeneracy label between two twist
fields in the same twisted sector. In addition, the coset construction restricts
$\lambda+\mu+\sigma$ to be even, and implies the field identification
$(\lambda,\mu,\sigma,\psi)\equiv(k-\lambda,\mu+h,\sigma,\psi)$. It is easy to
show that $\sigma=0$ corresponds to fields in the Ramond sector, while
those with $\sigma=1$ are in the NS sector.

Furthermore, one may compute the conformal dimension in the twisted sectors,
$$
\Delta_{(\lambda,\mu,\sigma,\psi)} =
\frac{c}{24} + \frac{(k-2\lambda)^2}{16h} \,,
\eqn\cdim
$$
if $\psi=+$, and one has to add $1/2$ if $\psi=-$. Here $c=3k/h$ is the central charge
of the minimal model, and $h=k+2$.

It is amusing to note that the conformal dimension \cdim\ is independent of $\sigma$
and is thus the same for R and NS sector fields. One may wonder how this is consistent 
with spacetime supersymmetry. To understand this, let us recall a few facts about 
the $\ZZ_2$ orbifold of the free boson/fermion system that represents the $S^1$ in the 
theory \central. In the twisted sector of (supersymmetric) $S^1/\ZZ_2$, fermions in 
the R sector are antiperiodic, while fermions in the NS sector are periodic. On the 
other hand, the bosons are always antiperiodic. Therefore, the contribution from the 
$S^1$ to the conformal dimension is $1/16$ in the R sector and $1/16+1/16$ in the NS 
sector. The difference is just the expected difference between R and NS sector conformal
weights for the supersymmetric compactification of 7 dimensions. Something similar 
happens in 8 dimensions. These considerations also show that there will be no tachyons 
from the twisted sectors.

Another important consequence of the formula \cdim\ is that there are
R ground state (which are characterized by $\Delta=c/24$) if and only if
$k$ is even and $\lambda=k/2$. In particular, when we combine the (orbifolds
of) $\N=2$ minimal model to form our $G_2$ holonomy CFT, there can be R
ground states in the twisted sector if and only if all levels of the minimal models
involved are even. 

The quintic model is the simplest example where there actually are no ground states
from twisted sectors. This Gepner model consists of 5 minimal models at level $3$.
It has $208$ RR ground states, in one-to-one correspondence with the elements of
the cohomology of the quintic. When we multiply with the two ground states of the 
boson/fermion system, and divide by the $\ZZ_2$ orbifold action, we are again left 
with $208=b_0+b_2+b_3+b_4+b_5+b_7=2(1+b_2+b_3)$ ground 
states\foot{The fact that
$b_1=0$ can be justified from a geometric perspective by the fact that the only 
nontrivial 1-cycle in the initial geometry
does not survive the $\Z_2$ projection.}. The geometry 
corresponding to the $\ZZ_2$ orbifold of the quintic Gepner model times $S^1$ is 
thus predicted to have
$$
b_2 + b_3 = 103 \,.
\eqn\pred
$$

\subsec{Geometric Interpretation}

We now turn to the geometric picture of the construction. We will focus on the quintic
Calabi-Yau, but some of our arguments will be more general. The quintic hypersurface
in $\P^4$ at the Gepner point is given by
$$
Q:=\left\{\sum_{i=1}^5 \left(Z^i\right)^5=0,~(Z^1,\dots,Z^5)\in\P^4\right\}\,.
\eqn{\defi}
$$
Quite generally, we expect the conformal field theories of the previous subsection to
describe geometries of the form $(CY_3\times S^1)/\Z_2$, where $CY_3$ is the Calabi-Yau
three-fold associated with the Gepner model \refs{\FKSS,\LYSC}. Here, the geometric action 
of $\Z_2$ is extracted from the CFT construction. More precisely, we will see that the 
$\Z_2$ acts by an antiholomorphic involution whose fixed point set is a special Lagrangian 
cycle in $CY_3$. In particular, the $G_2$ structure on $CY_3\times S^1$, which according 
to \HITC\ is specified by the 3-form
$$
\phi= {\rm Re}(\Omega)+ J\wedge \theta
\eqn{\struct}
$$
is left invariant. Here, $\Omega$ is the (unique) holomorphic 3-form on the Calabi-Yau
three-fold $CY_3$, $J$ is its K\"ahler form, while $\theta$ is the generator of
$H^1(S^1,\Z)$.

A candidate geometric $\Z_2$ action on $CY_3$ is best derived in the context of 
Landau-Ginzburg models, whose IR fixed points are the minimal model building blocks 
of the Gepner model. The important observation is that $\omega$ acts on the left- and
right-moving sectors of the CFT in the same way, namely by the simple sign flip of
the ${\rm U}(1)$ charge, see eq.\ \primauto. In the effective LG description, this is
reproduced by the action $\omega:~\Phi\mapsto\bar\Phi$, which is indeed a symmetry of 
the LG action. We note that the orbifold amounts to gauging the symmetry between
chiral and anti-chiral fields of the theory, and can not be confused with the
mirror symmetry of \HOVA, which exchanges chiral with twisted-chiral fields.
We conclude that the geometric action of $\omega$ on the Calabi-Yau is simply
complex conjugation of the coordinates, $Z^i\mapsto \bar Z^i$. It is then easy to
see that the $G_2$ structure {\struct} is preserved by $\omega$, since both $J$ and
$\theta$ change sign while ${\rm Re}(\Omega)$ is invariant.

It is worthwile pointing out that the identification of a geometric action of
$\omega$ need not be unique. 
For example, we can easily generalize it by twisting with automorphisms of the
Calabi-Yau space. For the quintic, this is achieved by including phases \BDLR\
that preserve the form of the equation \defi. More precisely, we can consider
the $\Z_2$ action
$$
\omega:~\Phi_i\mapsto \rho_i{\overline\Phi}_i~~{\rm 
with}~~\rho_i^5=1~~{\rm and}~~\prod\rho_i = 1\,.
\eqn{\geninv}
$$
For the given $G_2$ structure \struct, there are then $125$ choices of quintets 
$(\rho_1,\ldots,\rho_5)$. At the Gepner point, these choices are all equivalent 
by coordinate transformations, but not so away from it.

The fixed point loci of these involutions \geninv\ are special Lagrangian cycles,
described by the equations
$$
{\rm Im}((\rho_i)^{-1/2}Z^i)=0\,.
$$
Some properties of these cycles can easily be discovered by studying the 
restriction of the quintic equation to the fixed point set, which reads
$$
\sum_{i=1}^5 \left({\rm Re}((\rho_i)^{-1/2} Z^i)\right)^5=0\,.
$$

It is easy to see that this equation has unique real solutions for all choices
of $(\rho_1,\ldots,\rho_5)$. Furthermore, since the equation is projective,
the fixed point locus is $L=\Re\P^3$, whose fundamental group is
$$
\pi_1(\Re\P^3)=\Z_2\,.
$$
Since $\Re\P^3$ is orientable, Poincar\'e duality applies and we find that 
$$
H^1(L, \Z)=0 \,.
\eqn{\0}
$$
This last observation turns out to be very important in comparing the 
geometry with the CFT results.

Since the involutions \geninv\ that we wish to divide out have fixed points, we
must study the structure of the resulting space around those points and see
whether the singularities can be resolved. Because the $\Re\P^3$'s fixed by
\geninv\ are (special) Lagrangian cycles, the local structure of the quintic
around them is that of the total space of the cotangent bundle $T^*\Re\P^3$, and
$\omega$ acts on the cotangent directions.

The fixed point set of $\omega$ in $CY_3\times S^1$ actually consists of two copies
of the aforementioned $\Re\P^3$ cycles, because of the two fixed points on $S^1$.
Using open sets $\{U_i\}$ to cover the fixed cycles, the local geometry around
the fixed point loci is that of the $A_1$ singularity
$$
U_i\times (\Re^4/\Z_2) \,,
\eqn{\patch}
$$
where $\Z_2$ acts on $\Re^4$ by reversing the sign of all coordinates,
$x_i\mapsto -x_i$, for $i=1,\dots,4$.

Following Joyce \refs{\ja,\jb}, over each open set $U_i$  we can resolve the $A_1$ 
singularity in the normal bundle while preserving the natural $G_2$ structure 
\refs{\BR, \BRSA}\ that is invariant under the $\Z_2$ action,
$$
\phi=\gamma_1\wedge \delta_1+
\gamma_2\wedge \delta_2+ 
\gamma_3\wedge \delta_3 + \delta_1\wedge\delta_2\wedge\delta_3 \,.
\eqn{\localG}
$$
Here, the $\delta_i$ are constant orthonormal 1-forms on $U_i$ and $\gamma_i$
constant 2-forms on the resolved $\Re^4/\Z_2$. Thus, it is certainly possible
to resolve the singularities. The non-trivial question is whether the local $G_2$
structures can be glued together to a global one. The operation can be split into two
steps: first we reconstruct the geometry of $(T^*\Re\P^3\times \Re)/\Z_2$ around each
of the two disjoint components of the fixed point set and then reconstruct the full
manifold.

The resolutions of the singularities considered in refs.\ \refs{\ja,\jb} can be endowed
with a global $G_2$ structure. However, in our cases it is not hard to see that the
gluing fails to produce a globally-defined parallel three-form. The fundamental result
of \GR\ implies that we should look for a harmonic three-form. However, since 
$H^1(\Re\P^3,\Re)=0$ it follows that $\delta_i$ are exact 1-forms with support on 
$\Re\P^3$. This in turn implies that $\phi$ is exact and thus, using Hodge decomposition, 
cannot be a harmonic form. Thus, we conclude that the manifold resulting from the 
resolution of the singularities in the normal bundle of $\Re\P^3$ does not have a 
globally defined torsion free $G_2$ structure.

The fact that the singularities cannot be resolved while preserving the $G_2$
structure \struct\ is the geometric picture of the absence of twisted sector RR ground 
states in the CFT. We are thus really studying string theory on a singular manifold
of $G_2$ holonomy. But the CFT, which was constructed as an ordinary orbifold, is
completely non-singular. In particular, we would not expect any additional
non-perturbative massless states. This does not exclude the possibility that the CFT
can be deformed by exactly marginal operators that do not fall in supersymmetry 
multiplets and thus completely break spacetime supersymmetry. This is of course 
consistent with the results above since lack of $G_2$ structure prevents
the existence of parallel spinors. It would be interesting to systematically analyze
such possibilities.

Due to the presence of singularities, the geometry and topology of this space
are somewhat subtle to define. One can count the invariant forms as in \PAPI, but
this does not generally yield the correct dimension of the cohomology groups 
since one has to take into account additional contributions arising from the 
collapsed cycles. In our case, starting from the cohomology 
groups of the quintic and of the circle we find that there are no invariant 1-forms 
since $\omega$ acts on $S^1$ by reversing the sign of the coordinate and there are
no invariant 2-forms since the 2-form on the quintic changes sign under
the action of $\omega$. The invariant 3-forms one can construct are: the real part
of any element of $H^{2, 1}(Q)$, the real part of $\Omega$, and
$J\wedge\theta$ where $J$ is the K\"ahler form on the quintic and $\theta$
is the 1-form on the circle. We thus find 
$$
b_0=1 \qquad b_1=0 \qquad b_2=0 \qquad b_3=103
$$
From a geometric perspective it is not clear whether we should count any 
collapsed cycles. As discussed in Section 3, the CFT prediction for the
stringy Betti numbers is
$$
b_2^s+b_3^s=103\,,
$$
which suggests that for the $(Q\times S^1)/\Z_2$ the stringy Betti numbers 
count the $\Z_2$ invariant forms.

\newsec{Conclusions}

In this paper, we have presented examples of rational CFTs describing string
propagation on manifolds with $G_2$ holonomy. The chiral algebra of these models
was obtained as the fixed point algebra of the $\N=2$ superconformal algebra under
the mirror automorphism, extended by an appropriate spin $3/2$ field. Due to the
equivalence with $\ZZ_2$ orbifolds of Gepner models times a circle, the chiral
algebra contains the algebra that characterizes $G_2$ holonomy CFT as a subalgebra.

The construction is interesting for the conformal field theorist just as much
as for the geometer. We have mentioned the CFT aspects in detail in the text,
so let us here rather comment on some geometric implications.

One of the main results from CFT is that there are typically no twisted sector RR
ground states and supersymmetric moduli. We have argued for the quintic that this
reflects the fact that the orbifold cannot be resolved to a smooth $G_2$ holonomy
manifold. It is natural to expect other Gepner models to lead to a similar structure.
It may well be possible that the models obtained in this fashion are atypical,
and do not reflect generic properties of $G_2$ holonomy manifolds. Nevertheless, 
for the manifolds that can be constructed in this way, it follows that
all deformations of the manifold are actually inherited from
the Calabi-Yau, and hence there should be an interesting relation between the
respective moduli spaces. This should have interesting physical signatures.
In particular, due to the absence of twisted massless states, the ${\cal N}=2$ 
three-dimensional field theories obtained by compactifying string theory on
such manifolds are just {\it field theory orbifolds} of ${\cal N}=4$ theories.
Depending on the particular realization and by analogy with \BEJO,  this in
turn would imply that, in an appropriately chosen regime, some correlation
functions are the same as in the parent theory. It would be interesting to analyze
this in detail.

\ack{We would like to thank Samson Shatashvili for his inspiring lecture on
Exceptional Magic at the Les Houches 2001 Summer School. R.R.\ thanks Martin
Ro\v{c}ek for getting him interested in $G_2$ holonomy CFTs. We are grateful
to Lilia Anguelova, Harald Ita, and especially Christian R\"omelsberger for very
helpful discussions. The research of R.R. was supported in part by the DOE under 
Grant No.\ 91ER40618(3N) and by the National Science Foundation under Grant No.\ 
PHY00-98395(6T). The research of J.W.\ was supported in part by the National Science 
Foundation under Grant No.\ PHY99-07949.}

\listrefs
\end